\documentclass[useAMS,usenatbib]{mn2e}
\usepackage{natbib}
\usepackage{subfig}
\usepackage{graphicx}
\usepackage{multirow}
\usepackage{dcolumn}
\usepackage{rotating}
\usepackage{float}
\usepackage{hyperref}
\usepackage{amsmath}
\usepackage{epstopdf}

\newcommand{\Mpc}{$h^{-1}$ Mpc}
\newcommand{\kpc}{$h^{-1}$ kpc}
\newcommand{\Msun}{$h^{-1}$ M$_{\odot}$}
\newcommand{\Msunnh}{M$_{\odot}$}
\newcommand{\Mstar}{M$_{\star}$}
\newcommand{\Zsun}{Z$_{\odot}$}
\newcommand{\kms}{km s$^{-1}$}
\newcommand{\edisk}{$\epsilon_{\mathrm{disk}}$}
\newcommand{\ehalo}{$\epsilon_{\mathrm{halo}}$}
\newcommand{\ehaloo}{$\epsilon^{0}_{\mathrm{halo}}$}
\newcommand{\Mvir}{$\mathrm{M}_{\mathrm{vir}}$}
\newcommand{\Vvir}{$\mathrm{V}_{\mathrm{vir}}$}
\newcommand{\Vmax}{$\mathrm{V}_{\mathrm{max}}$}

\newcommand{\env}{environment}
\newcommand{\Env}{Environment}
\newcommand{\dmstar}{$\Delta \mathrm{m}_{\star}$}
\newcommand{\dmr}{$\Delta \mathrm{m}_{\mathrm{reheat}}$}
\newcommand{\dme}{$\Delta \mathrm{m}_{\mathrm{eject}}$}
\newcommand{\dms}{$\Delta \mathrm{m}_{\star}$}
\newcommand{\logten}{$\mathrm{log}_{10}$}

\begin{document}

\title[]{Thinking outside the halo: Tracing the large-scale distribution of diffuse cosmic metals with semi-analytic models}
\author[G. M. Shattow et al.]{Genevieve M. Shattow, Darren J. Croton, \& Antonio Bibiano \\
Centre for Astrophysics and Supercomputing, Swinburne University of Technology, Hawthorn, VIC 3122, Australia}

\maketitle

\begin{abstract}
With the installation of the Cosmic Origins Spectrograph on the \textit{Hubble Space Telescope}, measurements of the metal content of the low redshift intergalactic medium (IGM) are now available. Using a new grid-based model for diffuse gas coupled to the SAGE semi-analytic model of galaxy formation, we examine the impact of supernova feedback on the pollution of the IGM. We consider different assumptions for the reheating and ejection of gas by supernovae and their dependence on galaxy circular velocity and gas surface density. Where metals are present, we find the most likely metallicity to be $-1.5 < $\logten(Z/\Zsun)$< -1.0$ at $z = 0$, consistent with both observations and more sophisticated hydrodynamic simulations. Our model predicts that the regions of the IGM with the highest metallicities will be near galaxies with M$_{\star} \sim 10^{10.5}$\Msun\ and in environments of densities $\sim 10 \times$\ the mean. We also find that 90\% of IGM metals at $z = 0$ are ejected by galaxies with stellar masses less than $10^{10.33}$\Msun.
\end{abstract}


\section{Introduction}
\label{s:intro}

Diffuse gas fills the Universe, surrounding and connecting galaxies, the densest peaks of the cosmic web. This filamentary gas accounts for over half of the baryons at $z = 0$ \citep{Dave1999, Shull2012}, but is much less dense than the gas occupying galaxies and therefore more difficult to observe. Nevertheless, the structure of the intergalactic medium (IGM) has been observed in Ly-$\alpha$\ emission \citep{Cantalupo2014, Martin2014}, Ly-$\alpha$\ absorption \citep[etc.]{Lynds1971}, and X-rays \citep{Werner2008, Ma2009, Planck2013a}. Future instruments, such as the \textit{Square Kilometer Array}, will also be able to see the IGM in faint 21-cm emission \citep{Takeuchi2014}.

Cosmic gas is not necessarily pristine hydrogen and helium left over from the Big Bang. As galaxies evolve, they eject gas from their halos back into the diffuse regions of the cosmic web through various feedback processes, including supernovae and active galactic nuclei (AGN). The ejected gas includes metals created during the life cycle of stars, pushed into the IGM by outflows from galaxies. 

For example, using the Cosmic Origins Spectrograph on the \textit{Hubble Space Telescope}, \citet{Shull2014} have measured the metallicity of the IGM to be $\sim 0.1$\Zsun, a number in agreement with most theoretical estimates \citep[see e.g.][]{Cen2001, Fang2001, Chen2003, Oppenheimer2012}, using large number of sight lines.

Historically, modelling the cosmic gas in and in-between galaxies has been performed with hydrodynamic simulations \citep[e.g.][]{Popping2009, Dave2010}. Hydrosimulations are numerically detailed and accurate, yet computationally costly. Their detail makes them ideal for studying individual systems or small volumes where one may wish to explore the complexities of galaxy formation. Baryons can also be traced outside of halos, allowing modellers to study the IGM as well. Their cost, however, makes it difficult to run statistically significant numbers of simulations. This means they are prohibitive for exploring rare events, to sample unbiased distributions, or to demonstrate the statistics of a particular phenomenon. 

A complementary method is to run an N-body dark matter-only cosmological simulation first, then in post-processing couple an analytic model approximating the baryonic physics of galaxy formation \citep[e.g.][]{Croton2006}. This ``semi-analytic'' approach averages out many of the small-scale details, but produces integrated properties for galaxies and halos that can be directly compared against observed survey data. Semi-analytic models are computationally inexpensive to run and, for an equivalent run-time, can produce orders-of-magnitude more systems to study. Furthermore, the increase in speed allows large areas of parameter space to be explored.

\begin{figure} 
	\centering
	\subfloat{\includegraphics[width=3.3in]{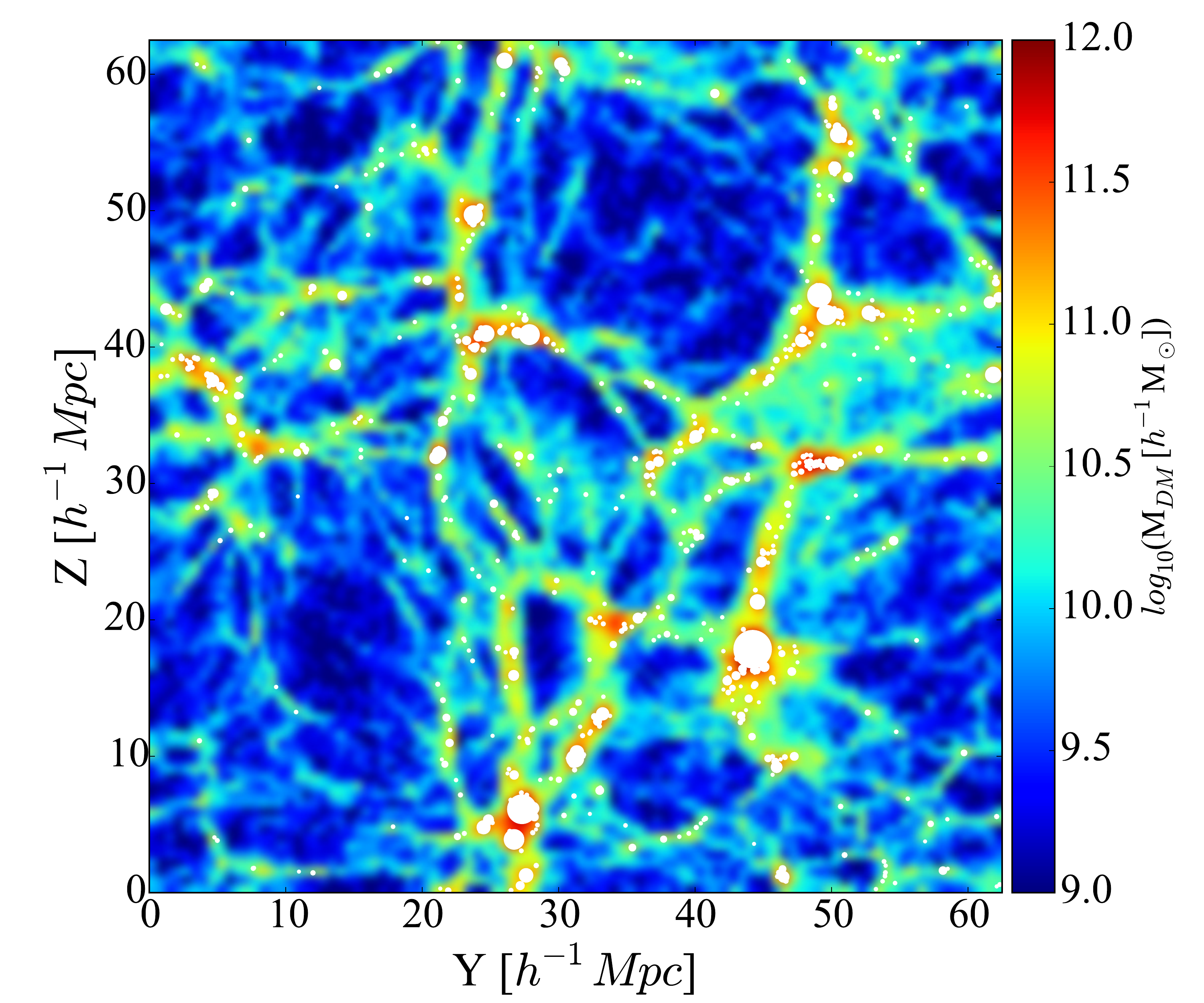}}
	\caption{Diffuse dark matter particles in a 1.25 \Mpc\ slice of milli-Millennium at $z = 0$. The white circles are the halos, scaled to their respective virial radii. \label{fig:slice}}
\end{figure}

Traditional semi-analytics follow the bound portions of the N-body simulation, treating the cosmic web only as it falls onto the galaxy. To look at the IGM, several extensions have been implemented using N-body simulations as a base. Most of these lack galaxy formation and are developed to address a single problem, such as creating mock HI and $^{3}$HeII observations \citep{Takeuchi2014}, finding the ionisation structure resulting from galaxies \citep[e.g][]{Kim2013, Park2014}, or measuring the light cone effects on the 21-cm signal \citep{Datta2014}.

We propose a generalized method for modelling the IGM that works in a similar fashion to traditional semi-analytic models. Rather than producing data focused only on a particular science question (e.g. ionisation, hydrogen density, etc.), our new model creates a full data set of gas in the intergalactic medium, which is coupled in a generic way with a semi-analytic galaxy formation model. We use the semi-analytic model SAGE \citep[][Croton et al., in prep]{Croton2006} to inject gas and metals into the IGM from galaxies through feedback, in addition to the primordial ``unprocessed" gas already in the IGM that we infer from the unbound dark matter in our N-body simulation. This combined IGM+galaxy model, called SAGE With All Matter Included (SWAMI), bridges the gap between semi-analytic models of galaxy formation and their more expensive cousins, hydrosimulations.

In this paper we introduce this new model and use it to consider the importance of supernova feedback on the metal content of the intergalactic medium. Using SWAMI, we test various scenarios of reheating and ejecting gas and metals from halos. 

This paper is organised as follows: in Section~\ref{s:methods}, we discuss the different components of SWAMI and the results of our fiducial model. In Section~\ref{s:varying}, we extend the fiducial model to test different reheating and ejection scenarios. Finally in Section~\ref{s:compare}, we discuss our results, how they compare to hydrosimulations and observations, and their implications for future observations. This is then summarised in Section~\ref{s:summary}.

\section{Modelling Galaxies and the Diffuse Cosmic Gas}
\label{s:methods}

\begin{figure*}
	\centering
	\begin{tabular}{cc}
		\subfloat{\includegraphics[width=3.3in]{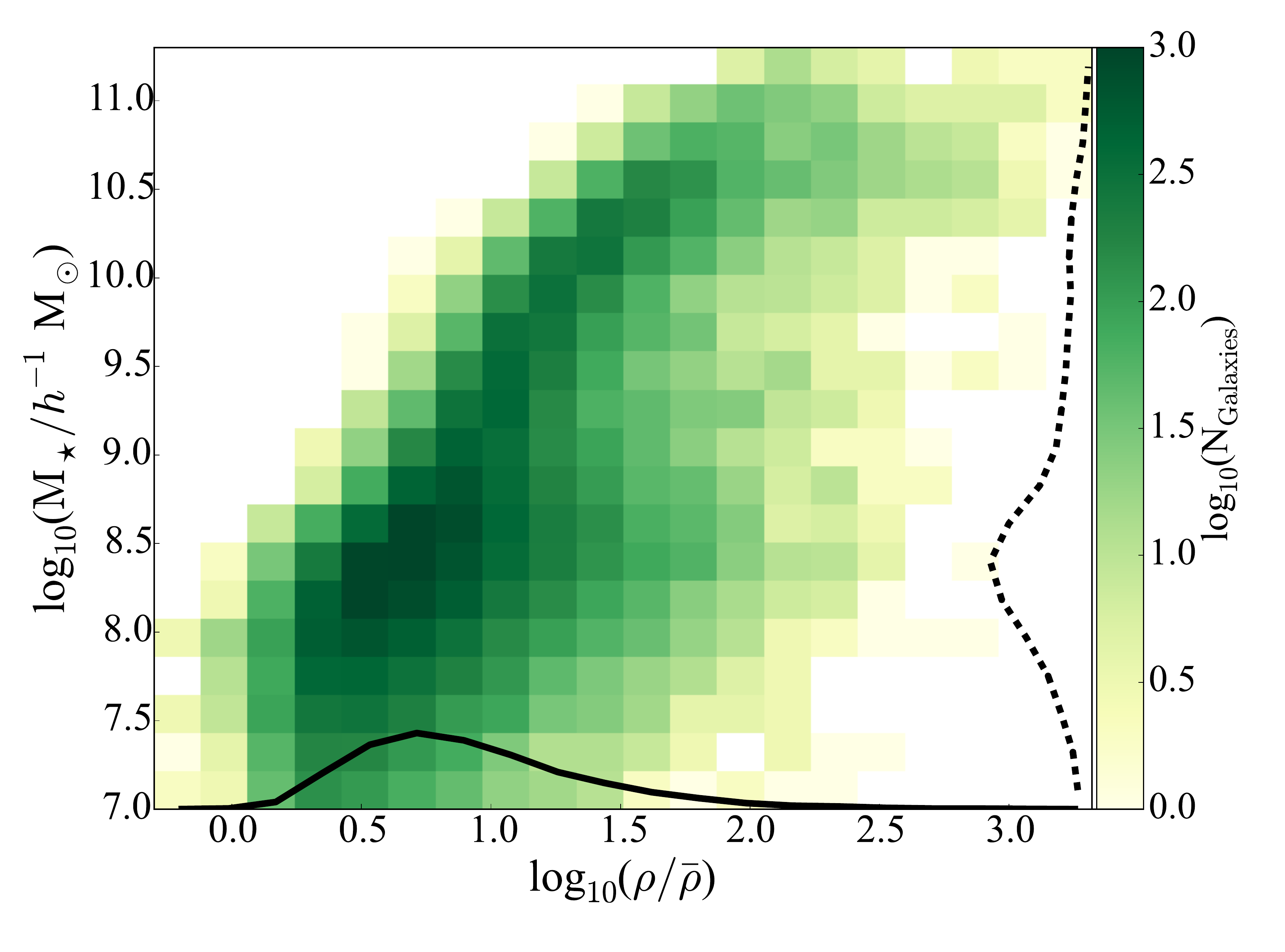}} &
		\subfloat{\includegraphics[width=3.3in]{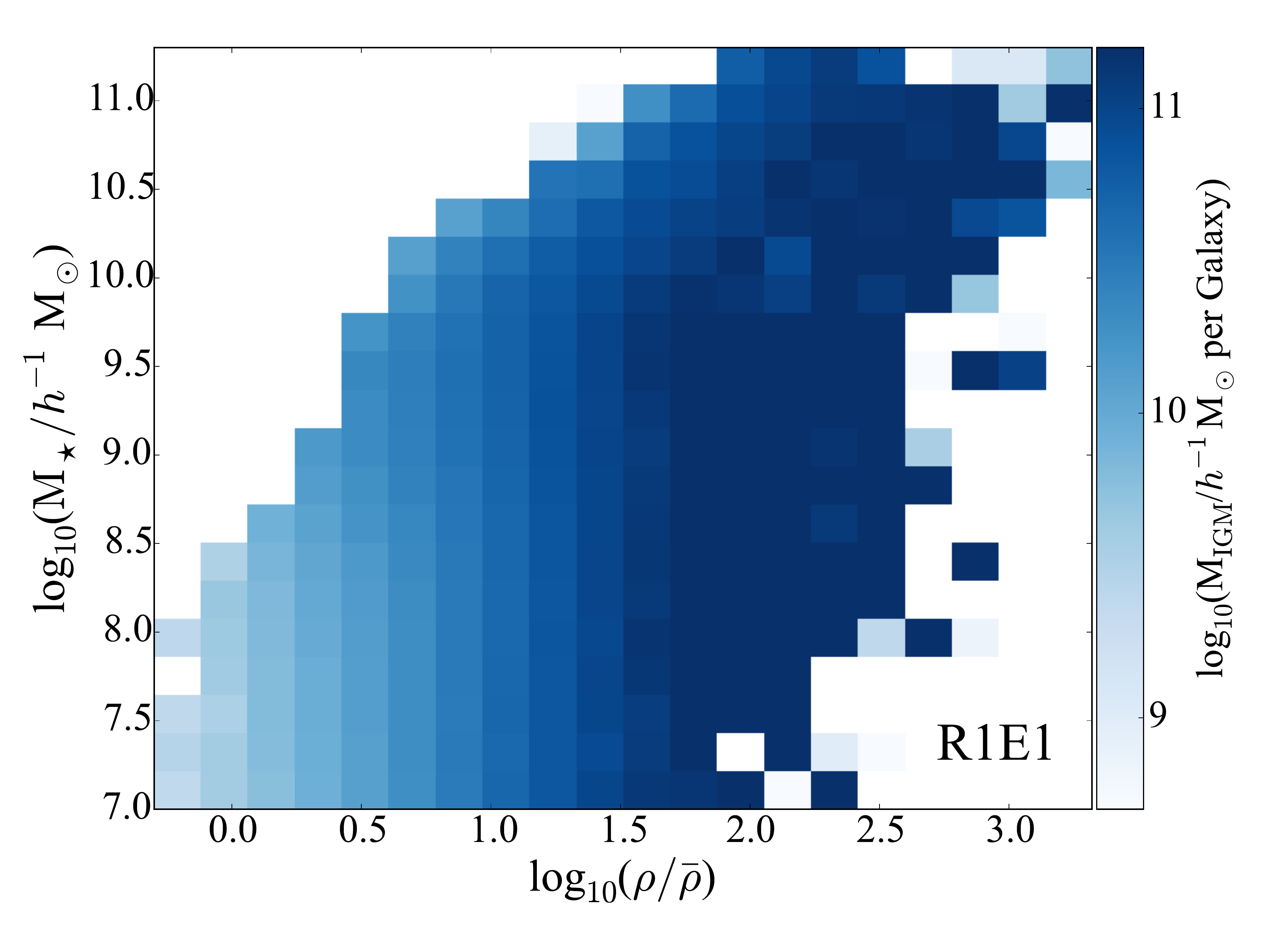}} \\
	\end{tabular}
	\caption{(left) Number of galaxies as a function of stellar mass and \env. The solid (dashed) line is a histogram of the \env\ (stellar mass) distribution. (right) Average mass of gas in the IGM in a cell as a function of stellar mass and \env. Both figures assume the fiducial SAGE feedback model.  \label{fig:mstars_IGM}}
\end{figure*}

Modelling the diffuse cosmic gas requires three separate stages. The first part is the dark matter only N-Body simulation, for which we use the milli-Millennium Run, a smaller version of the Millennium Simulation \citep{Springel2005a} developed for testing purposes\footnote{Full specs of the Millennium simulation can be found at \url{http://gavo.mpa-garching.mpg.de/Millennium/Help?page=simulation}.}, and described in Section~\ref{s:mMR}. 

The second part is the semi-analytic model of galaxy evolution we use to create galaxies within the dark matter halos of the N-Body simulation. For this, we use the Semi-Analytic Galaxy Evolution (SAGE) code \citep[][Croton et al., in prep]{Croton2006}, discussed more fully in Section~\ref{s:SAGE}.

The third part is a new semi-analytic model of the diffuse gas outside of halos, called SAGE With All Matter Included (SWAMI), using the unbound particles from the N-Body simulation. We discuss this part in Section~\ref{s:SWAMI}.

\subsection{milli-Milennium: The Dark Matter}
\label{s:mMR}

We use the smaller milli-Milennium rather than the full Millennium simulation because the particle data is available for the former, in addition to the halo data needed for the galaxy evolution model. The milli-Millennium simulation was run using the WMAP1 cosmological parameters \citep{Spergel2003} in a periodic box of side-length 62.5\Mpc, with a mass-per-particle of $8.6\times 10^{8}$\Msun.

We superimpose a $(100)^3$\ cell grid onto the box, giving each cell a side length of 625\kpc\ and an average of 19.68 particles (representing a total average mass per cell of $1.7\times 10^{11}$\Msun). Local density contrast ($\rho/\bar{\rho}$) is calculated for each cell using the number of dark matter particles located within, and this is used as our definition of \env. For reference, Milky Way-type systems at $z = 0$ contained entirely within a cell have a density contrast of about 6, not including any unbound dark matter also in the cell.

The particles are then divided into bound and unbound with a friends-of-friends plus gravitational unbinding procedure using the halo finder SUBFIND \citep{Springel2001}. The bound particles are assigned to halos, which are then used in SAGE to make galaxies. The unbound particles remain binned in the cells, which SWAMI uses to model the diffuse gas. An example of this is seen in Figure~\ref{fig:slice}, which shows a 1.25\Mpc\ (2 cell) deep slice of the milli-Millennium simulation. The white circles indicate halos (bound particles), scaled to their respective virial radii. The background structure shows the unbound particles binned into 625\kpc\ cells, which we assume traces the diffuse cosmic gas.

\subsection{SAGE: The Galaxy Model}
\label{s:SAGE}

Semi-analytic models have evolved considerably since they were first proposed by \citet{White1991}, although they still follow many of the same assumptions. Every halo hosts a galaxy and each galaxy is made up of several different baryonic reservoirs, including stars, cold gas (in the disk), hot gas (in the halo), and even ejected gas that has been pushed out of the halo. At each timestep, material cycles through the reservoirs according to analytical prescriptions of infall, gas cooling, star formation, galaxy mergers, and feedback processes. For our galaxy formation model we use SAGE, which is an update of the popular \citet{Croton2006} model, except for the modifications described below. 

For the creation and movement of metals, SAGE follows the prescriptions outlined in \citet{DeLucia2004}. Metals are created by supernovae and pushed into the disk at a rate proportional to the star formation rate, \dmstar: 

\begin{equation}
	\Delta \mathrm{m}_{\mathrm{ColdGas}}^{\mathrm{Z}} = \mathrm{Y}^\mathrm{Z}  \cdot \Delta \mathrm{m}_{\star}.
	\label{eqn:metals}
\end{equation}
We assume a yield, Y$^{\mathrm{Z}}$, of 2.5\%. As in many semi-analytic models, metal yield is a parameter of the model rather than an IMF-based calculation. Our value corresponds to a Salpeter initial mass function with an upper mass limit of 60M$_{\odot}$. In contrast, De Lucia et al. (2004) use a yield of 4.5\%, which is similar to a Chabrier IMF over the same mass range \citep{Madau2014}.

The energy output and timescales of the supernova parameters we use are based mostly on core collapse supernovae, which happen soon after star formation and are responsible for most of the alpha elements that are produced. Type Ia supernovae, in contrast, contribute to the heavy element population and occur on a delayed timescale from the star formation. Since elliptical galaxies tend to be older and therefore have had more of a build up of metals from Type Ia supernovae, the assumption of a constant yield directly proportional to the star formation rate does not hold as strongly. Our model predictions therefore are most robust with alpha elements produced in late type galaxies.

Once created, the metals then cycle through the various reservoirs -- stars, cold gas, hot halo gas, and ejected gas -- in a similar manner to the gas. Gas and metals can be ejected from a galaxy's halo by both supernova and quasars. Quasars affect the high mass end of the stellar mass function, whereas supernovae have more of an impact on the low mass end.

One of the main sources of uncertainty in semi-analytics (and understanding galaxy evolution in general) is the mechanism of feedback and its dependency on the evolving galaxy. To this end, we consider several different models for reheating and ejecting gas from the disk and halo with supernova winds. Below we describe our fiducial model. Later in Section \ref{s:varying} we consider several different models for reheating and ejecting gas from the galaxy and halo. These form a key part of our results with SWAMI.

\subsubsection{The Fiducial Galaxy Feedback Model}
\label{s:fiducial}

To set a baseline for our results we use the default SAGE model described in Croton et al. (in prep.), including all parameter choices. We refer the interested reader there for the more general details. Below we focus on the parts of the model most relevant to the topic of the current work, namely the modelling of supernova feedback and the movement of metals beyond the galactic disk into the halo and IGM.

Feedback processes, including supernovae and AGN, are crucial to the galaxy formation process. As stars form, a certain percent of the mass is automatically recycled back into the disk. The birth of very short-lived stars quickly result in supernovae that reheat the surrounding cold gas and push it out of the disk into the hot halo reservoir. We call this ``reheating'', and assume that the amount of gas that is reheated (\dmr) is proportional to the star formation rate and the disk mass loading factor, \edisk:

\begin{equation}
	\Delta \mathrm{m}_{\mathrm{reheat}} = \epsilon_{\mathrm{disk}} \cdot \Delta \mathrm{m}_{\star}.
	\label{eqn:massload}
\end{equation}
In our fiducial model, \edisk\ is a constant value with no dependence on the galaxy which hosts the supernova. This simple characterisation comes from early observations that measure \edisk$\sim 1-5$ \citep{Martin1999}.

The energy released during a supernova explosion is not completely consumed by the reheating of gas from the disk. It also drives winds that can eject hot gas from the halo. We assume conservation of energy, where the supernova energy left over from reheating the gas from the disk is used to kinetically drive gas from the halo into the IGM. Conservation of momentum has been ruled out as a likely mechanism for supernova ejection \citep{Murray2005}. 

From conservation of energy, the amount of energy available to eject gas, \dme, is
\begin{equation}
\Delta \mathrm{E}_{\mathrm{eject}} = \Delta \mathrm{E}_{\mathrm{SN}} - \Delta \mathrm{E}_{\mathrm{reheat}},
\label{eqn:Eeject}
\end{equation}
where
\begin{equation}
\Delta \mathrm{E}_{\mathrm{SN}} = \frac{1}{2}\epsilon_{\mathrm{halo}}^0 \Delta \mathrm{m}_{\star} \mathrm{V}_{\mathrm{SN}}^2
\label{eqn:Esn}
\end{equation}
and
\begin{equation}
\Delta \mathrm{E}_{\mathrm{reheat}} = \frac{1}{2} \Delta \mathrm{m}_{\mathrm{reheat}} \mathrm{V}_{\mathrm{vir}}^2.
\label{eqn:Ereheat}
\end{equation}
\ehaloo\ is the efficiency with which the supernova energy transforms into kinetically driven winds and V$_\mathrm{SN}$\ is the characteristic velocity of the wind, taken to be 680 \kms. We assume the reheated gas reaches virial equilibrium in the halo on a time-scale shorter than the time resolution of the simulation ($<$250-350 Myr), where the halo virial velocity is \Vvir. Assuming the escape velocity is $\sim$\ \Vvir, Equations \ref{eqn:massload} - \ref{eqn:Ereheat} combine to:
\begin{equation}
\Delta \mathrm{m}_{\mathrm{eject}} = \left( \epsilon_{\mathrm{halo}}^0 \left( \frac{\mathrm{V_{\mathrm{SN}}}}{\mathrm{V_{\mathrm{vir}}}}\right)^2 - \epsilon_{\mathrm{disk}}\right) \Delta \mathrm{m}_{\star}
\label{eqn:mej}
\end{equation}
We define the halo mass loading factor to be \ehalo $=$ \ehaloo $\left( \frac{\mathrm{V_{\mathrm{SN}}}}{\mathrm{V_{\mathrm{vir}}}}\right)^2$, which simplifies our expression for \dme\ to:
\begin{equation}
\Delta \mathrm{m}_{\mathrm{eject}} = (\epsilon_{\mathrm{halo}} - \epsilon_{\mathrm{disk}}) \cdot \Delta \mathrm{m}_{\star}.
\label{eqn:massloadh}
\end{equation}
As with the heating of the gas to the virial temperature, we assume ejection due to supernovae (and subsequent partial reincorporation) occurs within a timestep of the model.

After running the fiducial SAGE model on the halo merger trees of the milli-Millennium simulation we produce a catalogue containing 26,292 galaxies more massive than $10^7$\Msun\ at $z = 0$. In the left panel of Figure~\ref{fig:mstars_IGM} we show the distribution of these galaxies as a function of stellar mass and \env\ (local cell density contrast, as described in Section \ref{s:mMR}) Many galaxies fall below the stellar mass resolution limit of our model, which is taken to be M$_{\star} \approx 10^8$\Msun. Galaxies less massive than this usually have dark matter halos only just above the halo mass resolution limit of the simulation and are unlikely to have fully developed. Figure~\ref{fig:mstars_IGM} shows the expected trend of more massive galaxies living in denser \env s, with an additional grouping of low mass galaxies, just above the resolution, in \env s about $100\times$\ the mean density. The solid (dashed) line along the x-axis (y-axis) highlights the distribution of galaxies as a function of \env\ (stellar mass).

\subsection{SWAMI: The Diffuse Cosmic Gas Model}
\label{s:SWAMI}

The IGM in our diffuse gas model is considered in two parts - the baryons that trace the unbound dark matter particles from the simulation and the ejected gas reservoir of each galaxy modelled by SAGE. We assume that the unbound particles are traced with pristine gas and that all of the diffuse metals are in the immediate region of the galaxies that have ejected them. While this is in contrast to hydrosimulations which have revealed metals not only outside the halos of galaxies \citep[e.g.][]{Brook2012, Stinson2012}, but also in relatively empty regions of space \citep[e.g.][]{Oppenheimer2012}, the assumption that the gas that has not yet been processed through the halo is pristine is a reasonable approximation at low redshift. To address high redshift mixing, including a lower limit of the metallicity of infalling gas, a more complex model is needed, and will be the subject of future work. 

We start by overlaying an analytic mapping between cosmic baryons and the unbound particle grid of the simulation. We assign hydrogen and helium to the unbound matter according to the prescription in \citet{Madau2002}:
\begin{equation}
\bar{n}_{\mathrm{H}} = (\rho_{crit}/m_{\mathrm{H}})(1-Y)\Omega_b (1 + z)^3~,
\end{equation}
where $\rho_{crit}$, $m_\mathrm{H}$, $Y$, and $\Omega_b$\ are the critical density, mass of hydrogen, the primordial abundance of helium (0.24), and the baryonic density (0.04), respectively. This gives us a local hydrogen content of 
\begin{equation}
\begin{split}
\mathrm{M}_{\mathrm{H}} &= n_{\mathrm{H}}(x) \times \mathrm{v}_{\mathrm{cell}} = \bar{n}_{\mathrm{H}} (1 + \delta) \times \mathrm{v}_{\mathrm{cell}}  \\
&= 1.12 \times 10^{-5} \mathrm{cm}^{-3} \Omega_b h^2 (1 + \delta) \times \mathrm{v}_{\mathrm{cell}}~,
\end{split}
\end{equation}
where $\delta$\ is calculated using the total density of the cell, and v$_{\mathrm{cell}}$\ is the volume of the cell. The amount of diffuse gas is adjusted by subtracting off the mass of the galaxy. 

To partially couple the galaxy model to the diffuse model, we add the ejected reservoirs of gas for each fiducial SAGE galaxy to the IGM in the cell where each galaxy is located. The median amount of diffuse gas surrounding galaxies as a function of stellar mass and \env\ is shown in the right panel of Figure~\ref{fig:mstars_IGM}. As expected, the amount of IGM surrounding a galaxy is much more dependent on \env\ than on stellar mass, with IGM increasing with increasing density. The amount of IGM in a cell reaches its maximum at densities between 30 and 300 times the mean. Cells with higher densities than this contain mostly bound particles and thus do not have much IGM.

The ejected metals predicted by each SAGE galaxy and ejected due to supernova and AGN feedback are added to the cells in the same manner as the ejected gas. We assume perfect mixing within the cell but none outside. Previous studies have found that ejected matter is recycled through the halo and galaxy an average of three times over the life of the galaxy \citep{Oppenheimer2008}, and gas typically takes $\sim$ 1 Gyr to cycle through the galactic fountain \citep{Brook2014}, so our mixing model, so our mixing assumptions are a reasonable first-order approximation. Future work with a fully coupled model will be able to address the spread of metals out of the immediate region of each galaxy/halo system and follow the history of metals in the IGM as galaxies move between cells.

\section{Exploring Supernova Winds And Metal Ejection}
\label{s:varying}

The presence of metals in the hot galactic halo modulates the rate at which gas cools onto the galaxy, which in turn affects the star formation rate. By changing the mechanism by which metals and gas are injected into and ejected out of the halo, we alter not only the metal content of the IGM but also the galaxy life cycle itself. Thus, the model assumed for gas reheating and ejection will have consequences for the wider galaxy population, although the details remain ill understood. It is therefore worth exploring a range of possibilities and using these to compare with existing hydrosimulations and observations. In this section we focus on reheating (from the disk) and ejection (from the halo) separately, using modifications to the fiducial model disk and halo mass loading factors. We then examine the consequences of these modifications for selected galaxy and IGM properties.

\begin{figure} 
	\centering
	\subfloat{\includegraphics[width=3.3in]{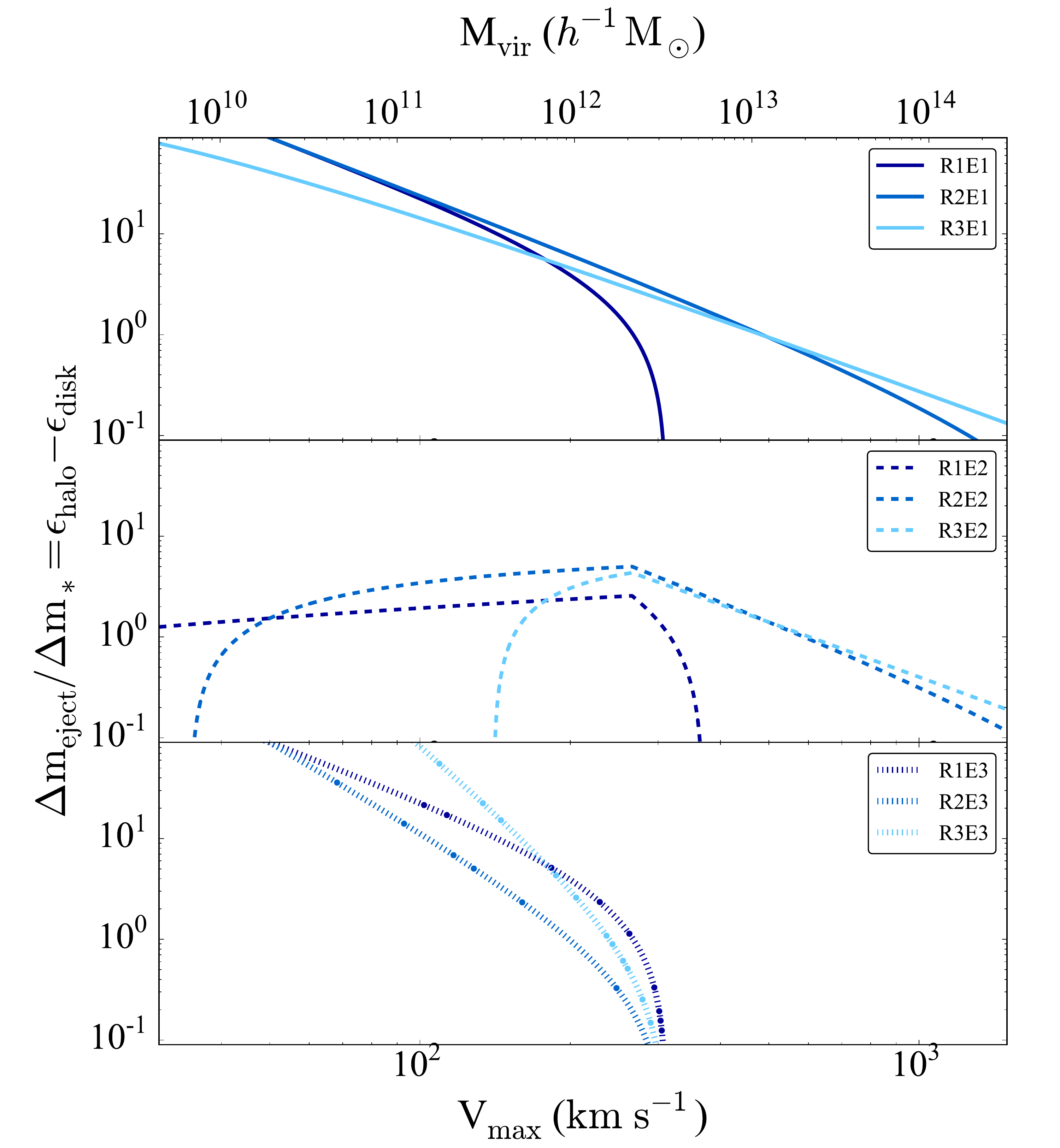}}
	\caption{The three panels show the nine combinations of heating and ejection models, used to calculate the ratio between ejected matter (\dme) and amount of new stars formed (\dms), as shown in Eqn. \ref{eqn:massloadh}. For all three panels, R1, R2, and R3 are the dark, medium, and light lines, respectively. E1, E2, and E3 are the solid, dashed, and dotted lines in the top, middle, and bottom panels, respectively.\label{fig:massload}}
\end{figure}

\subsection{Gas Reheating And Ejection Models}
\label{s:reheat}

To test the effect of different reheating assumptions on the metallicity of the IGM, we vary the disk mass loading factor, \edisk, according to various models suggested in the literature. We consider three in this paper: (R1) our fiducial model of a constant \edisk\ from \citet{Croton2006}, (R2) a disk circular velocity dependent \edisk\ as used by \citet{Dave2010}, and (R3) a cold gas surface density dependent \edisk\ from \citet{Lagos2013}:
\begin{equation}
\begin{array}{l l}
\epsilon_{\mathrm{disk}} = 3.0 & \quad \text{R1}\\
\epsilon_{\mathrm{disk}} = \frac{150\ \mathrm{km}\ \mathrm{s}^{-1}}{\mathrm{V}_{\mathrm{max}}} & \quad \text{R2} \\
\epsilon_{\mathrm{disk}} = \left(\frac{\Sigma_{\mathrm{ColdGas}}}{1.6\times 10^{15} \mathrm{M}_{\odot}\mathrm{Mpc}^{-2}}\right)^{-0.6} \cdot \left(\frac{f_{\mathrm{ColdGas}}}{0.12}\right)^{0.8} & \quad \text{R3} \\
\end{array} 
\label{eqn:edisk}
\end{equation}

\textbf{R1:} The simplest reheating model, R1 (our fiducial model), is commonly used by in semi-analytics and is motivated by the early work of \citet{Martin1999}, who showed that \edisk$\sim 1-5$ for galaxies over a wide range of circular velocities, albeit with a large scatter.

\textbf{R2:} Reheating model R2 is borrowed from a common implementation of galactic winds in hydrosimulations, such as by \citet{Oppenheimer2006} and \citet{Hirschmann2013}. A more physically motivated scaling than R1, recent observations have indicated the disk mass loading factor scales with circular velocity. For example, \citet{Bordoloi2014} find that the equivalent width of galactic outflows increases with stellar mass for galaxies of \Mstar$ < 10^{10.7}$\Msun\ at $z \sim 1$. We use halo \Vmax\ as a proxy for disk circular velocity in our modelling.

\textbf{R3:} Reheating model R3 is derived from dynamical modelling of superbubble expansion driven by supernovae \citep{Lagos2013}. Unlike R1 and R2, R3 takes its input from the SAGE galaxy properties directly (namely cold gas mass and stellar mass) and is therefore the most sensitive to the model assumptions and tunings (although note that we keep the same fiducial tuning for all results presented in this paper).

\vspace{5mm} 

As seen in Equation \ref{eqn:massloadh}, the pollution of the IGM from outflows is dependent on both \edisk\ and \ehalo. To that end, we also explore the halo mass loading factor, \ehalo. We test models that vary both the ejection efficiency \ehaloo, and the supernova velocity scaling. The three versions of \ehalo\ we use are:
\begin{equation}
\begin{array}{l l l}
\epsilon_{\mathrm{halo}} = \epsilon_{\mathrm{halo}}^0 \left(\frac{680\ \mathrm{km}\ \mathrm{s}^{-1}}{\mathrm{V}_{\mathrm{vir}}}\right)^2, & \epsilon_{\mathrm{halo}}^0 = 0.3 & \quad \text{E1}\\
\epsilon_{\mathrm{halo}} = \epsilon_{\mathrm{halo}}^0 \left(\frac{3 \times \mathrm{V}_{\mathrm{max}}}{\mathrm{V}_{\mathrm{vir}}}\right)^2, & \epsilon_{\mathrm{halo}}^0 = 0.3 & \quad \text{E2}\\
\epsilon_{\mathrm{halo}} = \epsilon_{\mathrm{halo}}^0 \left(\frac{680\ \mathrm{km}\ \mathrm{s}^{-1}}{\mathrm{V}_{\mathrm{vir}}}\right)^2, & \epsilon_{\mathrm{halo}}^0 = 0.1\ \epsilon_{\mathrm{disk}} & \quad \text{E3}\\
\end{array} 
\label{eqn:ehalo}
\end{equation}

\textbf{E1:} Ejection model E1 assumes a fixed value of \ehaloo, set at 0.3 (our fiducial model), which represents a constant efficiency of supernova energy being transformed into kinetic winds. It also assumes a constant supernova wind velocity.

\textbf{E2:} Ejection model E2 enhances the E1 by adding a dependence on the circular velocity of the galaxy. Here, we assume the supernova winds are not constant from one galaxy to another, but instead affected by the host galaxy. This is a similar assumption to our 2nd reheating model and a variation of the models used in e.g. \citet{Oppenheimer2006}, etc.

\textbf{E3:} Ejection model E3 adds a dependence on the disk mass loading factor through \ehaloo = 0.1 \edisk. This model comes from the assumption that the mechanism by which the supernovae reheat the gas from the disk is the same as the mechanism by which the gas is ejected from the halo. Physically, this model implies a strong interaction between the galaxy and its halo and that the amount of gas ejected depends only on the amount of gas that is reheated. Since hydrosimulations do not specifically eject gas from halos, this is a semi-analytic approximation for giving a particle a kick from a supernova and letting it travel.

\vspace{5mm} 

In Figure~\ref{fig:massload} we plot the behaviour of the nine reheating/ejection configurations. The three panels show the ratio of the mass of ejected gas to the mass of new stars formed during a timestep (equivalent to the difference between \ehalo\ and \edisk, Equation \ref{eqn:massloadh}), with top, middle, and bottom panels showing E1, E2, and E3, respectively. For all figures in this paper, dark, medium, and light coloured lines correspond to R1, R2, and R3, respectively. Likewise, solid, dashed, and dotted lines refer to E1, E2, and E3, respectively. R1E1 and R1E3 are identical, as the fiducial value of \ehaloo\ is 0.1$\times$ the fiducial value of \edisk. Since R3 is dependent on the cold gas and not directly on \Vmax\ or \Vvir, the R3 lines are from an empirical fit to the fiducial model and are meant for comparison only.

We use the same convention as \citet{Oppenheimer2008} and restrict E2 to a maximum \Vmax\ of 226 \kms. This corresponds to \Mvir $= 3\times 10^{12}$\Msun\ and limits the amount of matter that can be ejected from large halos. We include the factor of 3 so that galaxies at this maximum have the same supernova velocity as the other ejection models. There is a sharp cutoff for R1 with all ejection models at circular velocities corresponding to \Mvir $\simeq 10^{12}$\Msun.

\subsection{Stellar Mass Function}
\label{s:SMF}
\begin{figure*} 
	\centering
	\subfloat{\includegraphics[width=6.8in]{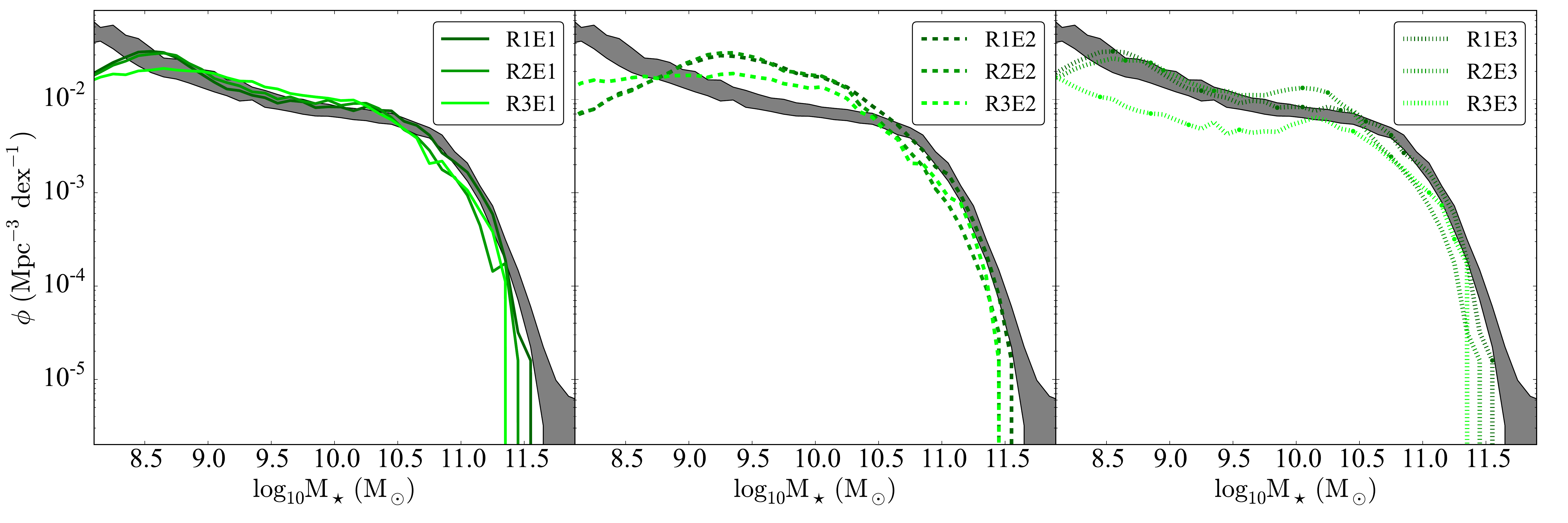}}
	\caption{The stellar mass functions of the nine models. Left to right, the panels are constant E1 (solid lines), E2 (dashed lines), and E3 (dotted lines). R1 (dark green lines), R2 (medium green lines), and R3 (light green lines) are shown in each panel. The shaded regions are the observed $z = 0$\ SMF from \citet{Baldry2008}. This figure uses $h = 0.72$\ to accurately compare observations and theoretical models which have different dependencies on $h$. \label{fig:SMF}}
\end{figure*}

Before examining the IGM, we start by quantifying the effects of our different reheating and ejection models on the galaxy population using the galaxy stellar mass function (SMF) at $z = 0$. Note that we do not retune SAGE between runs, only switch around the feedback model combinations. This provides a clean comparison of the impact each has on the model galaxy properties. The results are shown in Figure \ref{fig:SMF}, which, like Figure \ref{fig:massload}, has the three ejection models in three different panels. The green lines are the different models with dark, medium, and light green showing reheating models R1, R2, and R3, respectively. The solid, dashed, and dotted lines show ejection models E1, E2, and E3, respectively. The shaded region is the observed $z = 0$\ SMF from \citet{Baldry2008}.

Differences between the model combinations can mostly be seen at the low mass end of the SMF. For a constant ejection efficiency and supernova velocity, (E1, left panel), there is very little difference between the reheating model combinations, with all three matching the observations. On the other hand, galaxies that assume a circular velocity scaled ejection model (E2, middle panel) significantly overproduce the low mass population for all reheating models, with R3E2 sitting slightly closer to the observed SMF compared to R1 and R2. Ejection model E3 (far right panel) is the most dependent on the reheating model. Here, R1E3 is the same as R1E1 and is hence a good fit to the data, while R2E3 slightly over-predicts the SMF at $10^{10}$\Msunnh\ and under-predicts it below. R3E3 sits below the observed SMF at all masses.

None of the feedback model combinations have a significant effect on the high mass end of the stellar mass function. This is to be expected, as star formation at the high mass end is typically suppressed by AGN feedback, which we are not varying. The high mass end is slightly under-predicted for all models because we are using the milli-Millennium simulation, which does not have a significant number of high mass galaxies.

The nine combinations that are shown in Figure \ref{fig:massload} can be directly related to the stellar mass functions in Figure \ref{fig:SMF}. Gas that has been ejected must first fall back onto the halo before it can re-cool and settle onto the disk and form stars. Therefore, models that eject less gas at lower masses (e.g. E2, middle panel) will produce more galaxies in that same mass range. Likewise, R3E3 ejects the most gas at low masses, so it produces the fewest low mass galaxies. An empirical fit to the fiducial model has \edisk(R3) $\propto$ \Vmax$^{-2.24}$, making it more than an order of magnitude larger than \edisk(R2) ($\propto$ \Vmax$^{-1}$) for sub-L$^*$ galaxies.

Interestingly, R3 produces measurably fewer low mass galaxies than the other models do when the supernova wind speed is dependent on the galaxy's circular velocity (E2), but still more than observations show, despite ejecting the least amount of gas in the mass range of interest (light blue line in the middle panel of Figure \ref{fig:massload}). This is due to \edisk\ being very large at low masses and the supernova velocity being dependent on the circular velocity of the galaxy (and therefore the total available energy being much lower than in E1). Most of the energy is therefore expended on reheating the gas from the disk, leaving both less energy to eject the gas and less cold gas in the disk to form stars right away. The reheated gas does not make it out of the halo, but it still has to cool back onto the disk to make stars, suppressing the SMF relative to the other reheating models.

\subsection{Total Metals in the IGM}
\subsubsection{... vs. Stellar Mass}
\label{s:MZvMstars}

Figure \ref{fig:MZ_mstars} shows the total mass of the metals ejected into the IGM by all galaxies of a certain mass. As in Figure \ref{fig:SMF}, the dark, medium, and light coloured lines are reheating models R1, R2, and R3, respectively. The solid, dashed, and dotted lines are ejection models E1, E2, and E3, respectively. 

For all models, the net ejected metals peaks at roughly M$_{\star} \approx 10^{10}$\Msun. For the fiducial ejection model, E1, our fiducial and velocity-scaled reheating models, R1 and R2, peak at slightly above $10^{10}$\Msun, but the cold gas scaled reheating model, R3, is slightly below. Reheating model R3 produces fewer metals in the IGM than the other reheating models with the same ejection, while R1 and R2 produce similar amounts of metals. R1E2 and R2E2 differ by a factor of 1.5, but are still considerably closer to each other than the are to R3E2. R1E3 is not shown because it is the same as R1E1. Despite these differences, for all models at $z = 0$, 90$\%$ of the metals in the IGM have been ejected by galaxies of \Mstar $< 10^{10.33}$\Msun.

Sub-L* galaxies are responsible for most of the metals ejected into the IGM, and the amount of metals produced and ejected are proportional to the star formation rate (Equations \ref{eqn:metals}\ and \ref{eqn:massloadh}). It would, therefore, stand to reason that the models that produce the most sub-L* galaxies would eject the most metals into the IGM. For example, for our reheating dependent ejection model (E3), the superbubble derived reheating method (R3) produces the least amount of stars (right panel, Figure \ref{fig:SMF}) and ejects a considerably lower amount of metals than our fiducial model.

We find a different situation with the circular velocity scaled supernova wind ejection model (E2). All three reheating models produce an excess of sub-L* galaxies (middle panel, Figure \ref{fig:SMF}), indicating a high \dms, but have both the lowest and highest amounts of ejected metals (dashed lines, Figure \ref{fig:MZ_mstars}). The variation in metal ejection for E2 models is due to the other part of the ejection equation (Equation \ref{eqn:massloadh}), \ehalo\ - \edisk. The middle panel of Figure \ref{fig:massload}\ shows that, for the mass range in question, \ehalo-\edisk\ for E2 is highest when coupled with R2 and lowest when coupled with R3. This is reflected in the scalings in Figure \ref{fig:MZ_mstars}.

Similarly, our fixed halo ejection efficiency and fixed supernova velocity ejection model (E1) produces the same amount of stars regardless of the reheating model (left panel, Figure \ref{fig:SMF}), indicating that \dms\ is roughly constant, so any variation in the amount of ejected metals relies entirely on the \ehalo - \edisk. The top panel of Figure \ref{fig:massload}\ shows the superbubble driven reheating model (R3) is slightly suppressed at sub-L* masses compared to the other reheating models, so the R3E1 model (light solid line, Figure \ref{fig:MZ_mstars}) puts the least amount of metals into the IGM for that ejection model.

\subsubsection{... vs. Environment}
\label{s:MZvrho}

Figure \ref{fig:MZ_rho}\ shows the total metals in the IGM as a function of the \env. As described in Section \ref{s:mMR}, \env\ is the density contrast of the dark matter particles in the cell. Like the above figures, the dark, medium, and light coloured lines are reheating models R1, R2, and R3, respectively. The solid, dashed, and dotted lines are ejection models E1, E2, and E3, respectively. 

Our results show that the mass of metals ejected into the IGM peaks in \env s between 10 and $30 \times$\ the mean density, which are the typical \env s of small groups. The ejection model E2 tends to peak at the lower end of this \env\ range ($\sim 10 \times \bar{\rho}$) for all three reheating models. E1 and E3 peak in the middle ($\sim 18 \times \bar{\rho}$) for R1 and R2, and at a higher density for R3E3. 90\% of ejected metals are found in \env s less dense than $40 \times \bar{\rho}$. 

\begin{figure} 
	\centering
	\subfloat{\includegraphics[width=3.3in]{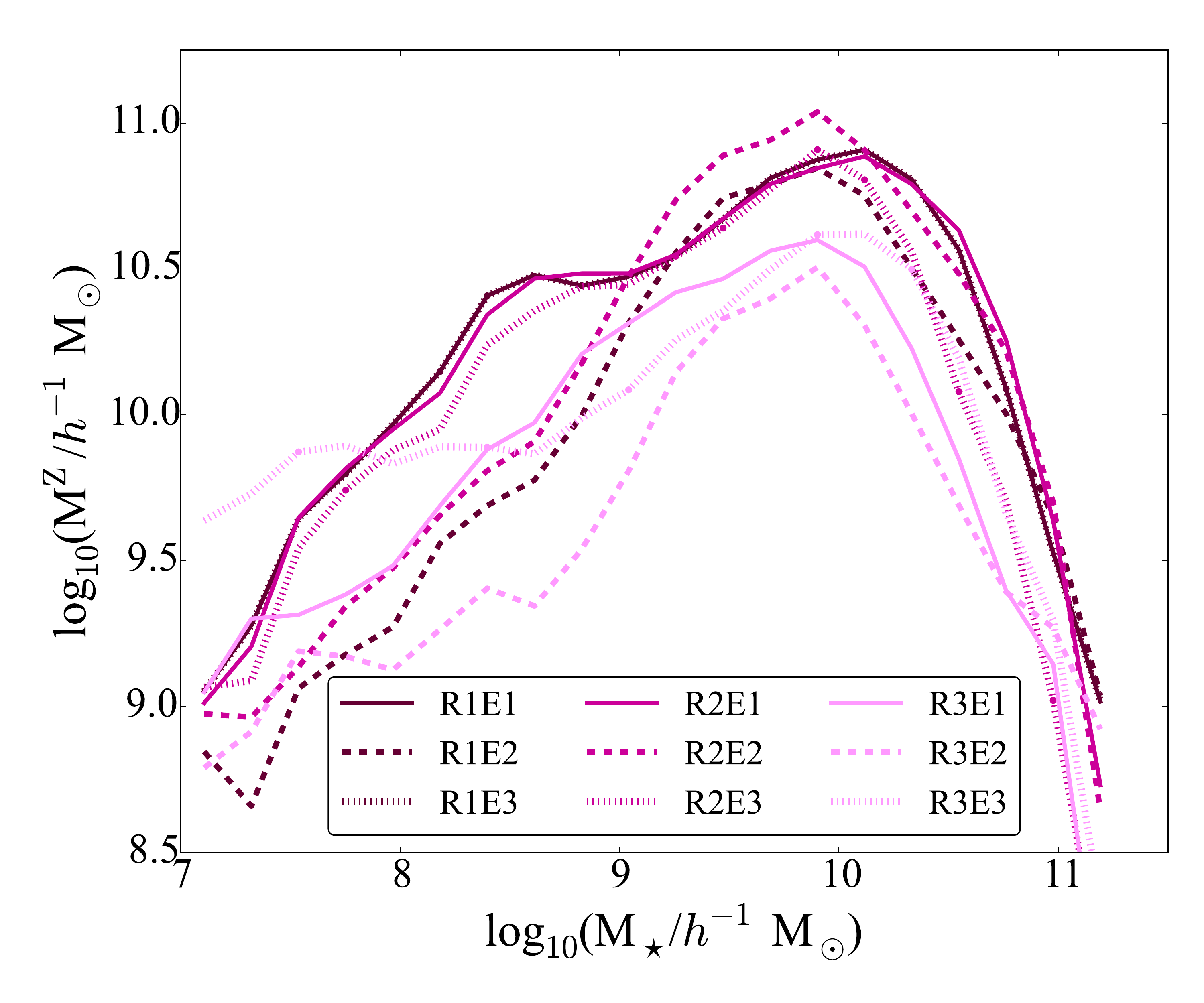}}
	\caption{Total ejected metals as a function of stellar mass. The dark/medium/light pink lines are reheating models 1, 2, and 3, respectively. The solid/dashed/dotted lines are ejection models 1, 2, and 3, respectively. \label{fig:MZ_mstars}}
\end{figure}

\begin{figure} 
	\centering
	\subfloat{\includegraphics[width=3.3in]{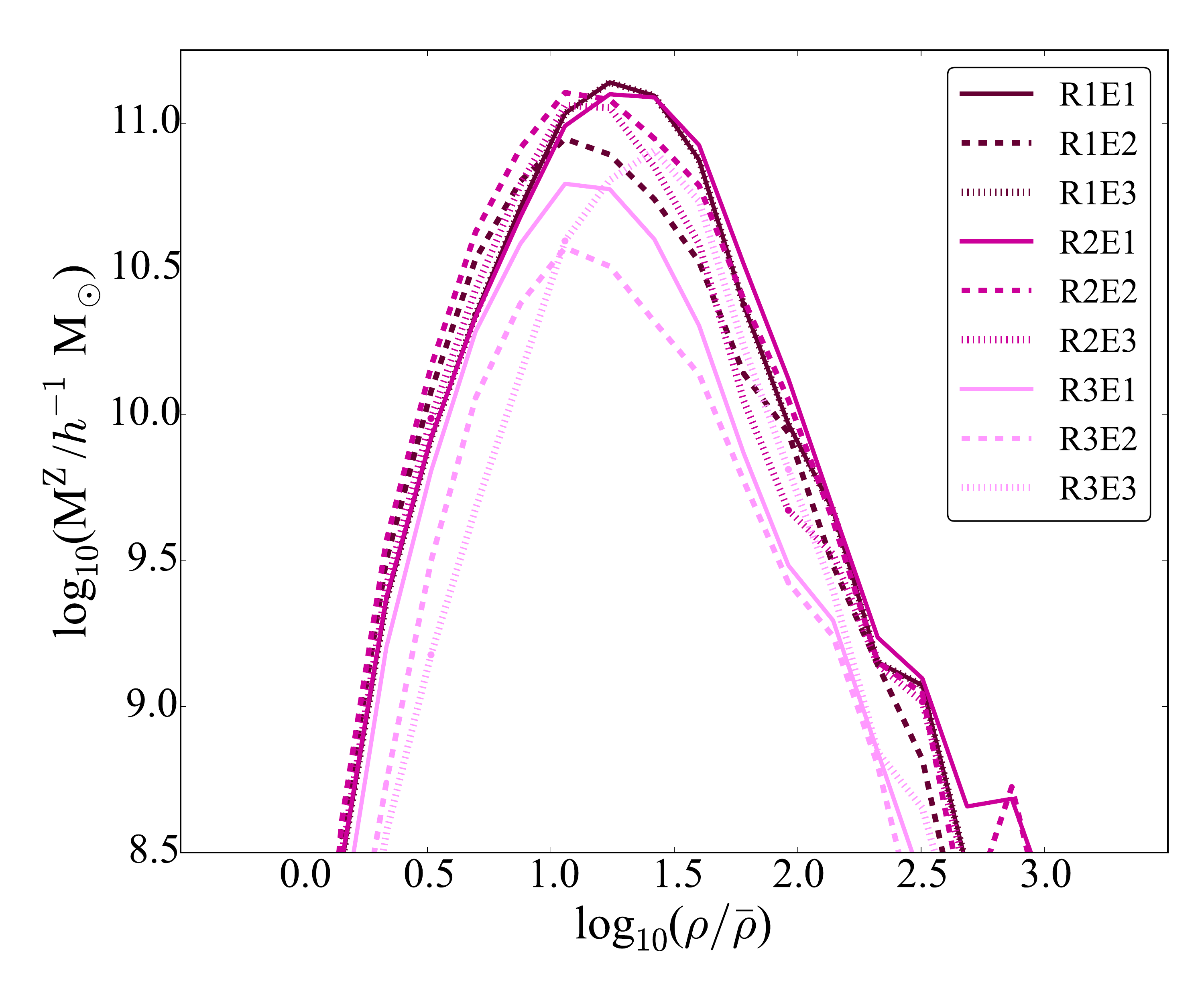}}
	\caption{Total ejected metals as a function of \env. The dark/medium/light pink lines are reheating models 1, 2, and 3, respectively. The solid/dashed/dotted lines are ejection models 1, 2, and 3, respectively. \label{fig:MZ_rho}}
\end{figure}

Galaxy mass scales roughly with density (Figure \ref{fig:mstars_IGM}), with larger galaxies evolving in denser \env s. Nevertheless, \env\ is much more closely related to halo mass than stellar mass. Comparing models as a function of environment is akin to comparing them as a function of halo mass. The models show very little variation in their slopes at lower densities, especially compared to the scatter between the models at low stellar masses in Figure \ref{fig:MZ_mstars}. There is more variation between the models at high densities, but their shapes are remarkably similar, differing only in their scaling.

\Env, and by extension halo mass, is more important for finding where the ejected metals are (in regions between 10 and $40 \times \bar{\rho}$). The distribution of densities where metals are found in the IGM, i.e. the shape of the distribution, is not model dependent, and the bulk of IGM metals will be found near small groups. The biggest difference between the models is in the net amount of ejected metals, and because of this scaling, a given \env\ can have a vastly different metal content, depending on the model.

\begin{figure}
	\centering
	\begin{tabular}{c}
		\subfloat{\includegraphics[width=3.3in]{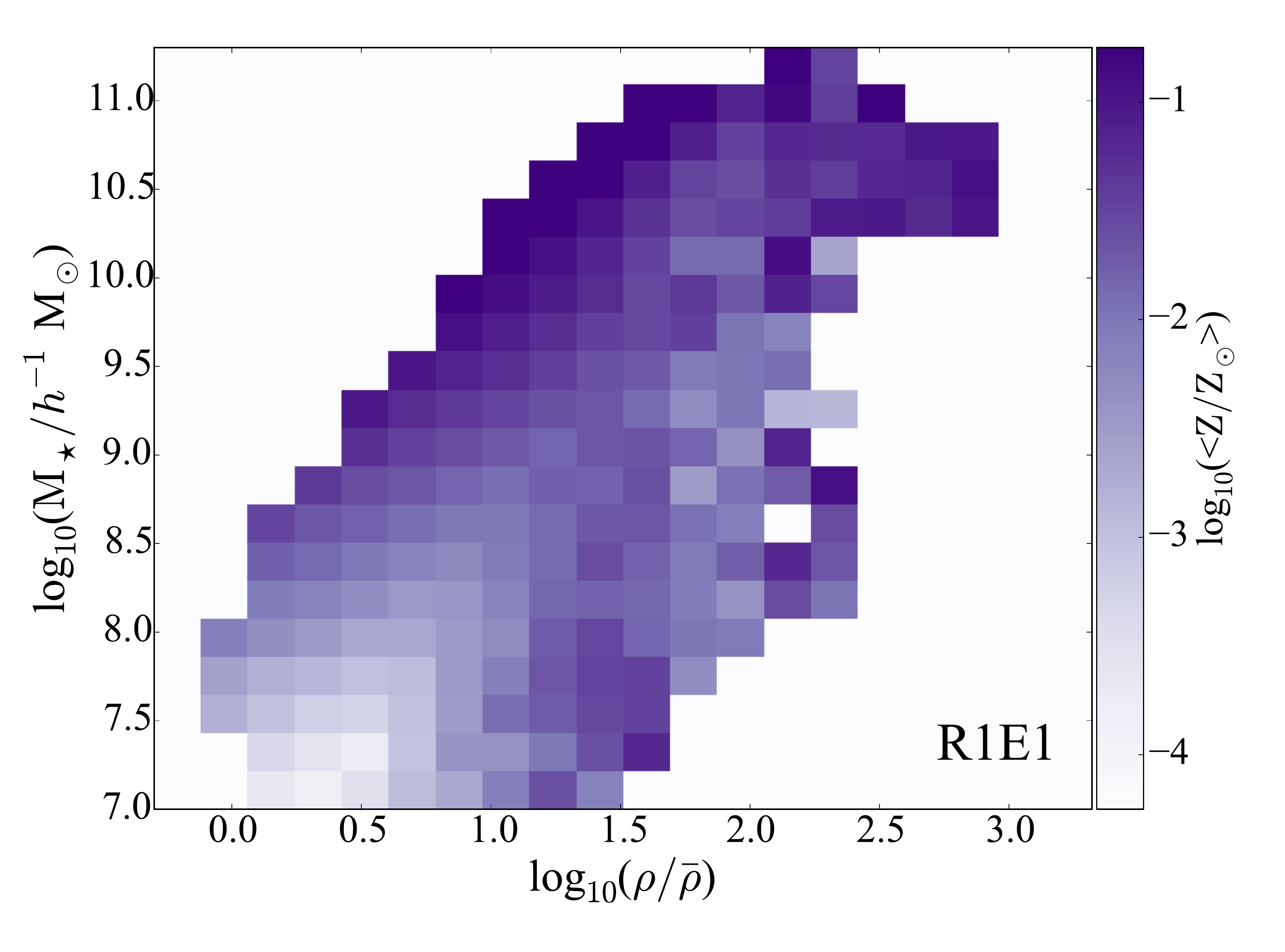}} \\
		\subfloat{\includegraphics[width=3.3in]{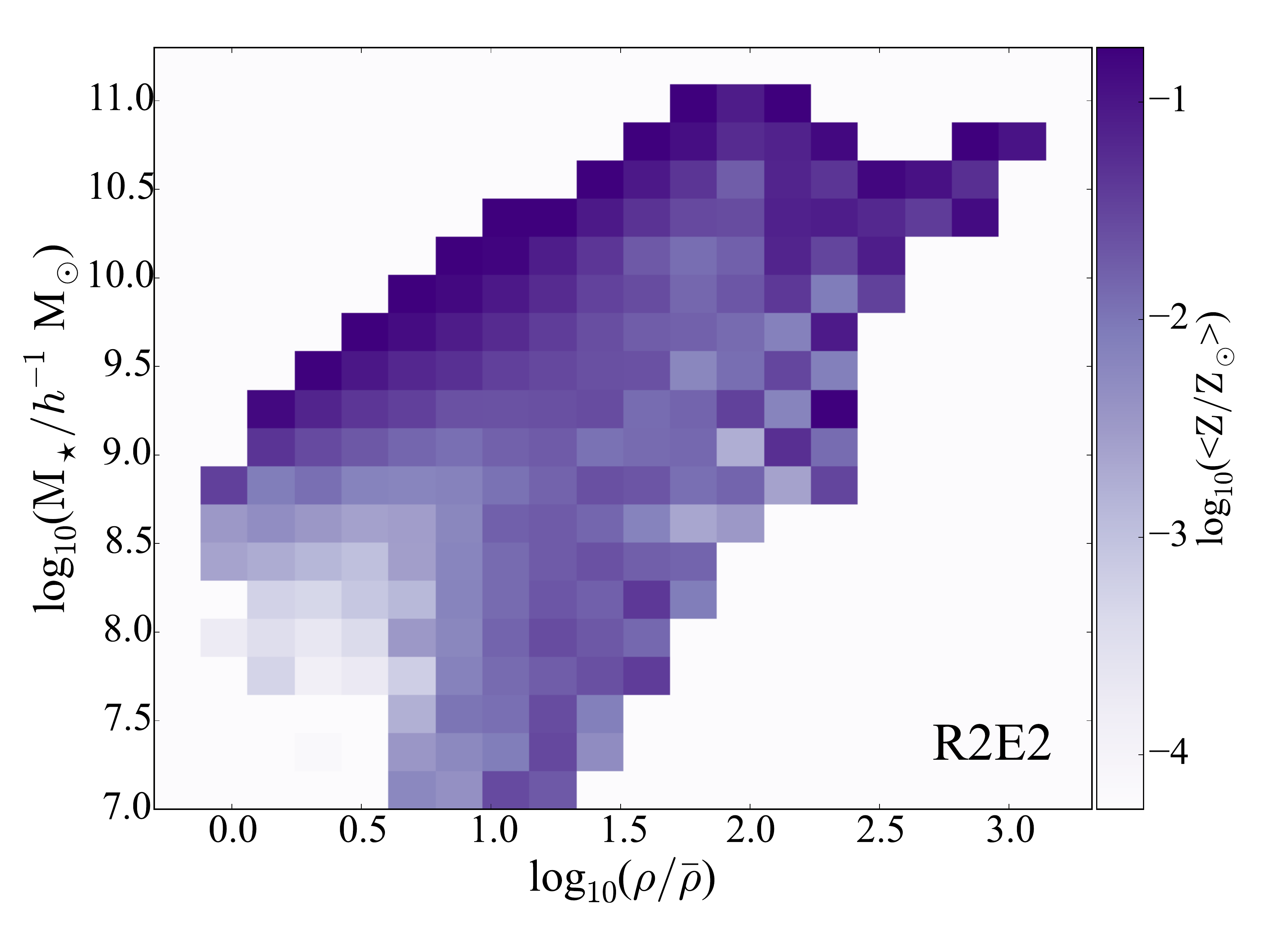}} \\
		\subfloat{\includegraphics[width=3.3in]{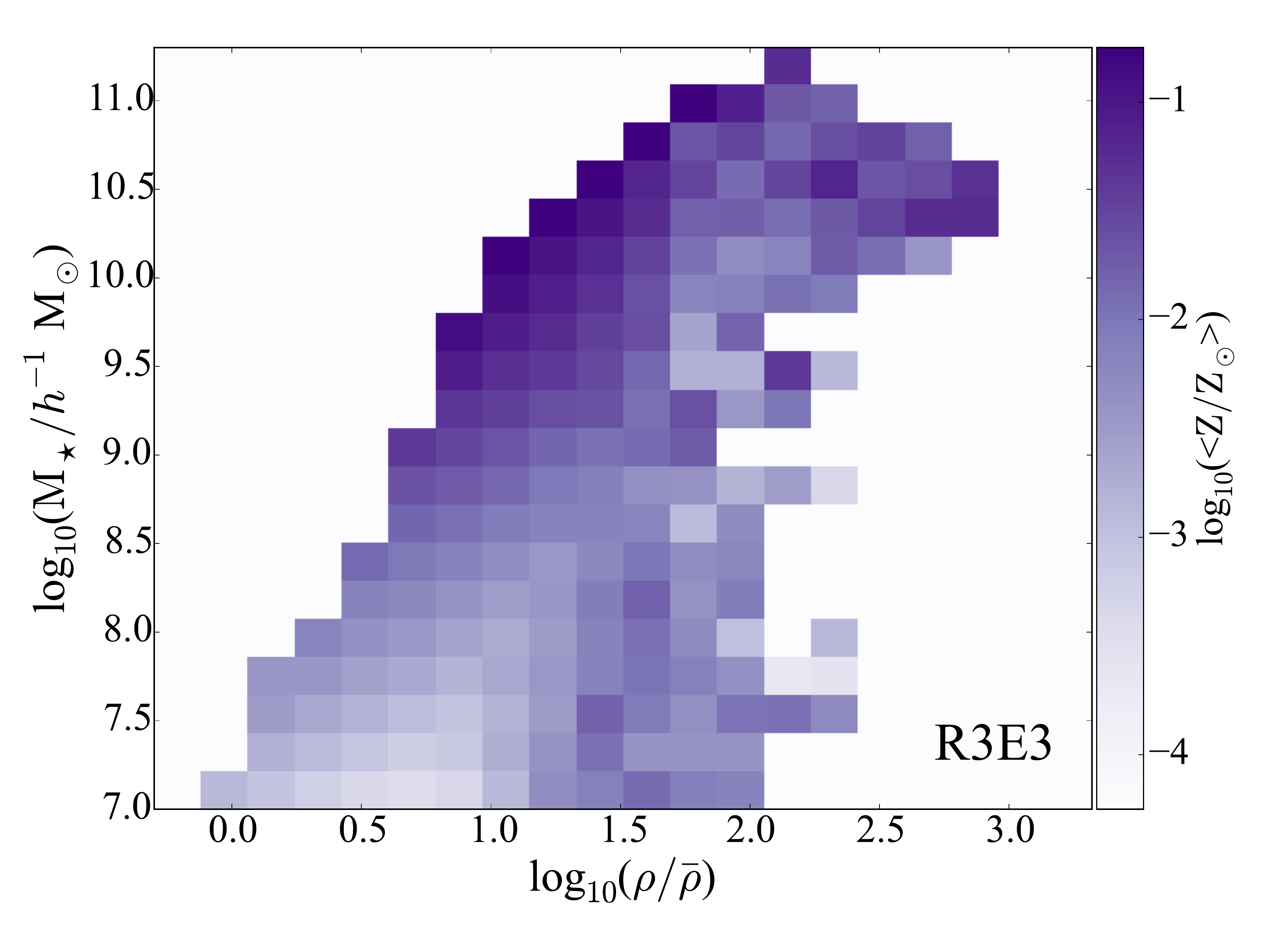}} \\
	\end{tabular}
	\caption{Average metallicity of the IGM around a galaxy as a function of stellar mass and \env. Top to bottom, the panels are R1E1, R2E2, and R3E3, respectively. The distributions are very similar for constant ejection models. \label{fig:ZZsun}}
\end{figure}


Total metal mass in the IGM is more dependent on ejection models than on reheating models. For all ejection models, R1 and R2 have similar total mass and the same peak density. R3 has a much lower total mass than R1 and R2 for a given ejection model. For a given reheating model, the various ejection models are much less tightly grouped, with no strong pattern emerging.

\subsection{Ejected Gas Metallicity}

Figure \ref{fig:ZZsun}\ shows the average metallicity of the IGM as a function of stellar mass and \env\ for three of the models (R1E1, R2E2, and R3E3), showing bins with at least 5 galaxies. This figure is analogous to Figure \ref{fig:mstars_IGM}, but is now coloured by average IGM metallicity in the bin rather than number of galaxies or average surrounding IGM gas mass. The average metallicity peaks at slightly sub-L* galaxies ($10.0 <$ \logten(\Mstar/\Msun) $< 10.5$) in \env s $\sim 10 \times$\ the average density, with R2E2 peaking at slightly lower densities and stellar masses. The metallicity distributions of the models that are not shown closely resemble the distribution of the same ejection model.

Overall, the average metallicity in a stellar mass/density bin ranges from 0.1\Zsun\ to below 1.0e-4\Zsun. Again, ejection models have a greater effect on the distribution of metallicities than the reheating models, which is unsurprising, as the reheating portion of SAGE moves gas from the disk to the halo and the ejection portion moves it out to the IGM. 

The metallicities also show a second peak for very low mass galaxies in similar density \env s. These galaxies are below the resolution of the code, so this is most likely a numerical issue. Satellite galaxies do not have ejected reservoirs, as they are added to the host halo during infall, and therefore are not included in this figure.

Focusing specifically on the mass and density range where the most metals are ejected, we define a subset of galaxies with $9 \le$\ \logten(\Mstar/\Msun) $\le 11$\ and $0.5 \le$ \logten($\rho/\bar{\rho}$) $\le 2.0$. By limiting the density, we eliminate the regions of space that are not dense enough to have measurable clouds of HI as well as higher density regions that preferentially host large groups and clusters where the hot halo dominates the gas content. Our stellar mass lower limit allows us to include only very well resolved galaxies. The high end cut is to remove galaxies larger than L*, which likely live in halos that are too massive to allow metals to escape. Galaxies in this sample account for 86\% of the metals in the IGM in the fiducial model and 85-95\% of galaxies in all of the other models. These numbers can be found in Table \ref{tab:subset}.

\begin{figure}
	\centering
	\subfloat{\includegraphics[width=3.3in]{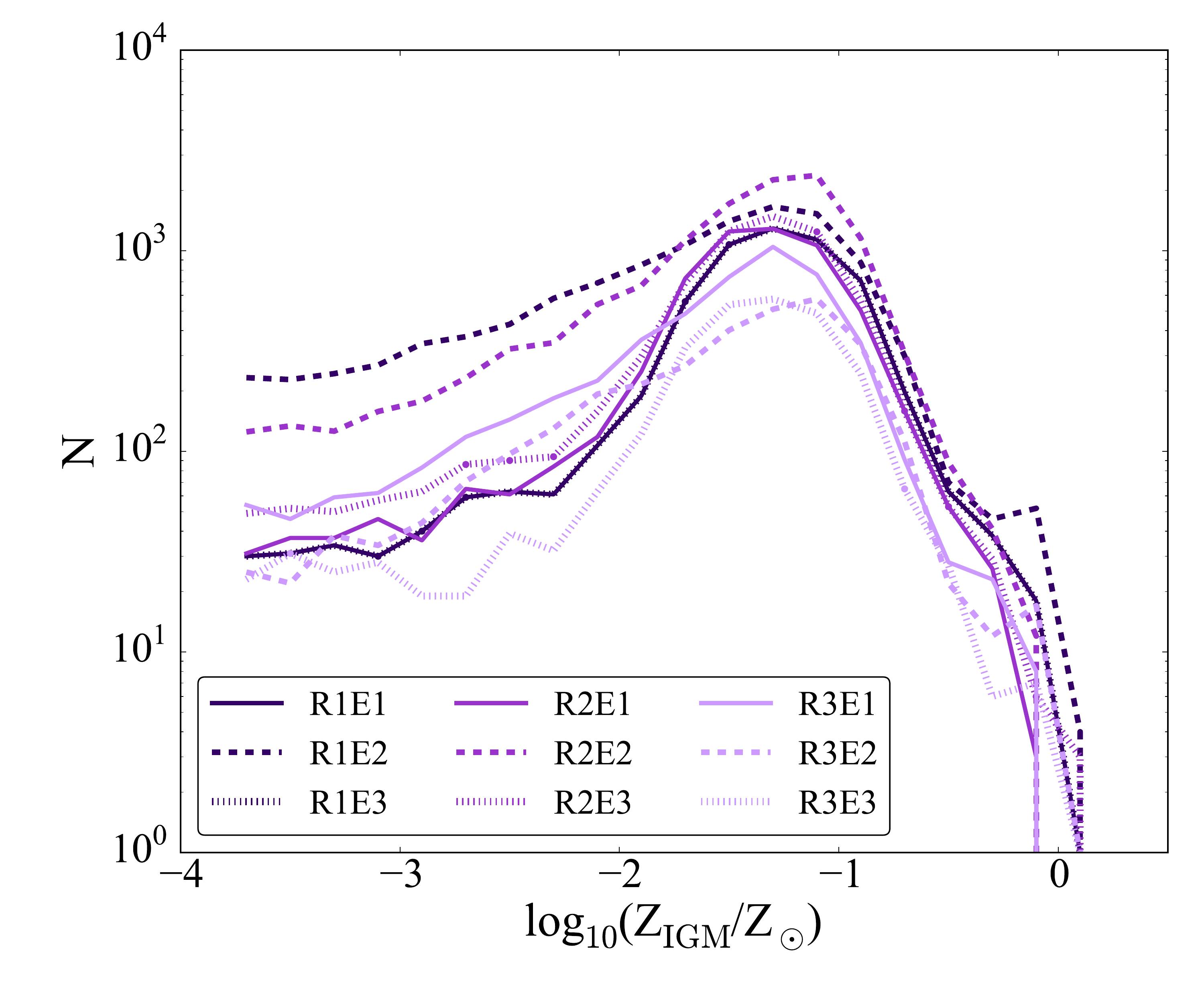}}
	\caption{Distribution of metallicities in the IGM surrounding galaxies in regions of density $0.5 \le$ \logten($\rho/\bar{\rho}$) $\le 2.0$\ and with stellar mass $9 \le$\ \logten(\Mstar/\Msun) $\le {11}$. The dark/medium/light lines are reheating models 1, 2, and 3, respectively. The solid/dashed/dotted lines are ejection models 1, 2, and 3, respectively. \label{fig:Z_hist}}
	
\end{figure}

Figure \ref{fig:Z_hist}\ shows the distribution of metallicity measurements in the IGM surrounding galaxies in our subset. The models vary in both the value of their most likely metallicity (the peak of the distribution) as well as the number of low metallicity regions surrounding galaxies in the subset. The variation in scaling comes from the relative number of galaxies in this range, which can be seen in the SMFs in Figure \ref{fig:SMF}.

The most likely metallicity (i.e. the peak in the distribution) for each of the models is between $-1.5 <$ \logten(Z/\Zsun) $ < -1.0$. E2 models have slightly higher peak metallicities for a given reheating model. Despite ejecting fewer metals in general, the metals they do eject go into less dense regions (see Figure \ref{fig:MZ_rho}), decreasing the amount of IGM in the cell and increasing the metallicity.

R1E2 and R2E2 have more low metallicity IGM regions than other models. This is partially due to the excess of galaxies produced by E2 models (middle panel, Figure \ref{fig:SMF}) - a factor of 2 in this subset. These extra galaxies are below the mass threshold in the other models, and inhabit environments on the lower cusp of our density range. They eject small amounts of metal into lower density regions, doubling the number of regions in the IGM where metallicity can be measured, but having little effect ($\sim 10 \%$) on the relative amount of IGM metals found in the subset. So few of the galaxies in R3E2 eject metals that the increase in number of galaxies produced is entirely countered by the low percentage of metal ejecting galaxies.

The ejection models not only affect the amount of gas being ejected but also if gas is ejected at all. In models that use R1, 90-93\% of galaxies in this stellar mass and density range eject gas and metals, and in R2 models, the range is 89-92\%. In R3 models, however, the range is 35-87\%; here, the presence of ejected gas is more tied to the reheating model than the ejection model. The percentages are listed in Table \ref{tab:subset}. The relative numbers of galaxies that have ejected metals are apparent in Figure \ref{fig:Z_hist}. 

\begin{table}
  \centering
  \begin{tabular}{| l | ccc | ccc |}
  	\hline
  	   &  \multicolumn{3}{c|}{\% of Gals in Subset} & \multicolumn{3}{c|}{\% of IGM Metals} \\
  	   &  \multicolumn{3}{c|}{w/ Ejected Metals} & \multicolumn{3}{c|}{in Subset} \\
  	     	   & R1     &   R2   &    R3     & R1     &   R2   &    R3  \\ 
  	   \cline{2-7}
  	E1 & 93\% & 92\% & 70\% & 86\% & 86\% & 85\%                \\
  	E2 & 90\% & 92\% & 35\% & 93\% & 95\% & 94\%               \\
  	E3 & 93\% & 89\% & 87\% & 86\% & 85\% & 85\%               \\ \hline
  \end{tabular}
  \caption{Columns 1-3: Percentage of galaxies with densities in the range $0.5 <$ \logten($\rho/\bar{\rho}$) $< 2.0$\ and having a mass within $9.0 < M_{\star}/$\Msun $< 11.0$\ that have ejected metal reservoirs. Columns 4-6: The percentage of total metals that are found in the IGM surrounding galaxies in this subset.}
  \label{tab:subset}
\end{table}

Since we do not change any parameters in SAGE other than the mass loading factors, we cannot rule out any of the models from just the stellar mass function and supernova feedback models.

\section{Comparisons to Other Work}
\label{s:compare}

\begin{table}
	\centering
	\begin{tabular}{| l | ccc | ccc |}
		\hline
		& \multicolumn{3}{c|}{\% of Metals Found} & \multicolumn{3}{c|}{$\Omega_{\mathrm{Z}}\times 10^{-5}$} \\
		& \multicolumn{3}{c|}{in the IGM} & \multicolumn{3}{c|}{in the IGM} \\
		& R1     &   R2   &    R3     & R1     &   R2   &    R3    \\ \cline{2-7}
		E1 & 8.1\% & 9.5\% & 4.6\% & 0.835 & 0.866 & 0.383 \\
		E2 & 4.6\% & 8.0\% & 2.6\% & 0.422 & 0.390 & 0.201 \\
		E3 & 8.1\% & 7.4\% & 6.8\% & 0.835 & 0.727 & 0.420 \\ \hline
	\end{tabular}
	\caption{Columns 1-3: The percentage of total metals that are found in the IGM. Columns 4-6: The total density of metals in the IGM.}
	\label{tab:Ztable}
\end{table}

\subsection{SWAMI vs. Hydrosimulations}
\label{s:hydro}

\begin{figure} 
	\centering
	\subfloat{\includegraphics[width=3.3in]{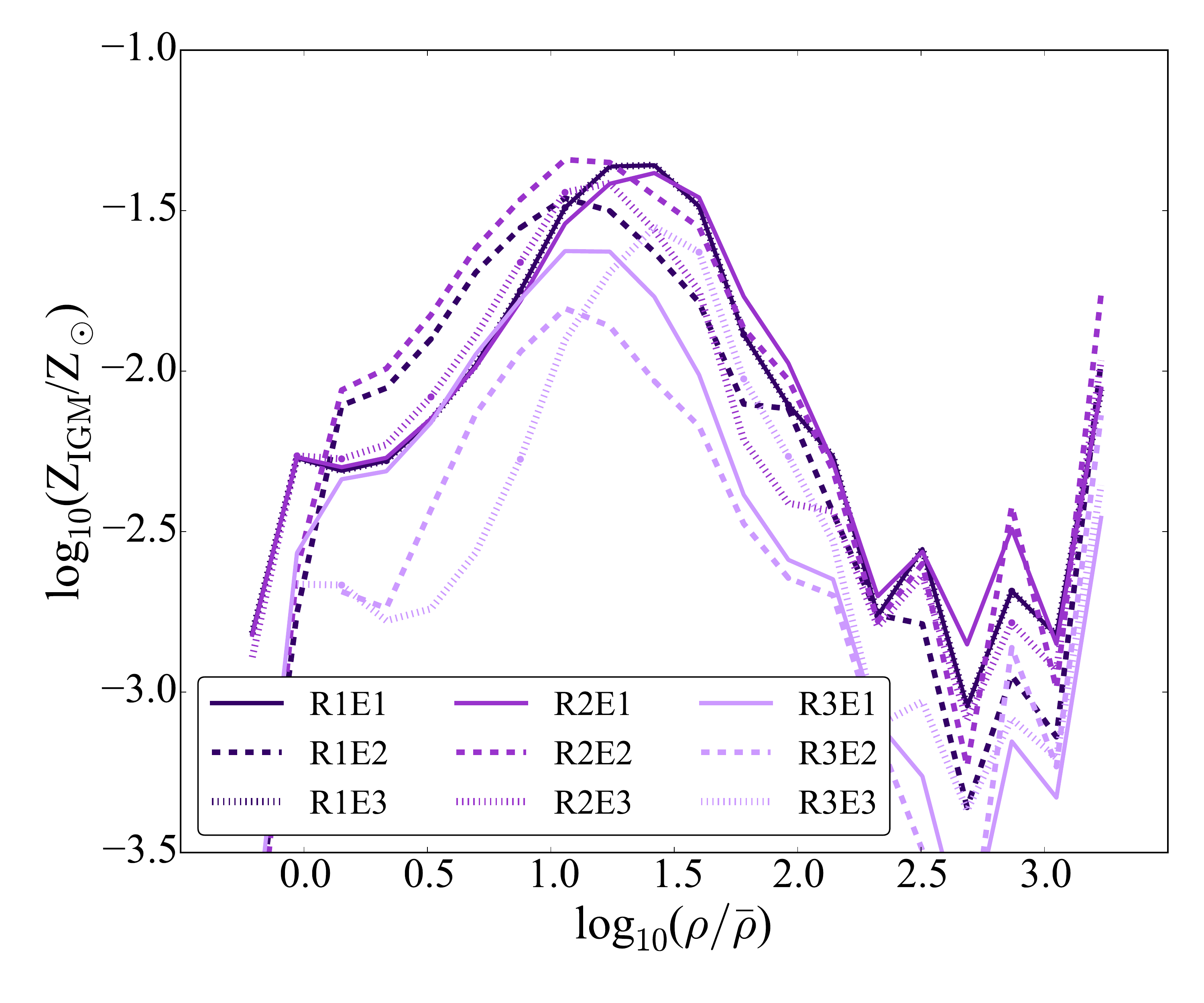}}
	\caption{Total metallicity as a function of \env s. The dark/medium/light lines are reheating models 1, 2, and 3, respectively. The solid/dashed/dotted lines are ejection models 1, 2, and 3, respectively. \label{fig:Z_rho}}
\end{figure}

Theoretical models of the intergalactic medium tend to be relegated to hydrosimulations, so it is natural to begin by comparing our results to those. In Figure \ref{fig:Z_rho} we show total metallicity versus \env. By total metallicity, we mean the total mass of metals in a certain \env, divided by the total IGM gas mass in the same \env. It is comparable to Figure 4 in \citet{Oppenheimer2012}, who also find $-1.5 <$ \logten(Z/\Zsun) $< -1.0$\ at $\rho \sim 10 \times \bar{\rho}$\ in all of their models with wind.

We do not find the same high metallicity in high density \env s that \citet{Oppenheimer2012} do for all of their models. This is mostly due to their inclusion of hot halo gas in their analysis. Larger galaxies reside in denser \env s (Figure \ref{fig:mstars_IGM}) and in larger halos. Larger halos have more hot gas and a higher metallicity than the IGM in general because metals merely have to be pushed out of the disk rather than out of the halo entirely.

The upturn we see at $\rho > 300 \times \bar{\rho}$\ is due to the lack of ``pristine" IGM at high densities. Most of the dark matter particles in these \env s are bound in halos, so the IGM has the metallicity of the ejected gas, rather than the metallicity of the mixed ejected+pristine gas found at lower densities.

At low densities, where halos tend to be smaller and therefore the hot halo is less of a concern, we find the metallicity drops off sharply. This is a good match to \cite{Oppenheimer2012}'s preferred model (vzw) of momentum conserved winds. Their constant wind (cw) model, which is similar to our R1 models, show a flat distribution in the metallicity in the range of $\bar{\rho} < \rho < 10 \times \bar{\rho}$, and a sharp decrease at lower densities. 

\citet{Stinson2012} and \citet{Brook2012} also show metals outside the halo, with a sharp drop off at a few hundred kpc. Their OIV column density profiles indicate that the presence of metals is more of a smooth gradient through the halo into the IGM than our model is able to currently convey. We hope to explore this further in future work.

\subsection{SWAMI vs. Observations}
\label{s:obs}
One of the many questions about metals in the Universe is whether they are all accounted for by current observations \citep{Pettini1999} or is there a percentage that we are ``missing". This is a difficult question to answer due to uncertainty in the amount of metals that have been produced since the Big Bang \citep{Madau2014} and the difficulty in measuring the quantity of metals in the IGM and hot halo gas. While theoretical models have the option of considering the IGM and its metals in any region of interest, current observational studies are limited to absorption lines. This makes one-to-one comparisons difficult, although the number of measurements is increasing with recent instruments allowing for more statistical studies. 

We can use the SWAMI data to compare the net amount of metals in the IGM as well as the distribution of measured metallicities. Both observations \citep{Shull2014} and hydrosimulations \citep{Oppenheimer2012} predict 10$\pm$5\% of metals produced are in the IGM at $z = 0$. We find that in 5 of our 8 unique models, M$^{\mathrm{Z}}$(IGM)/M$^{\mathrm{Z}}$(Total) $\sim$ 5-10\%. Of the other three, two more (R1E2 and R3E1) are just below $5\%$, leaving only R3E2 (where most of the supernova energy is used to heat the disk gas) with much too low a metal content. The values for each model can be found in Table \ref{tab:Ztable}.

\subsubsection{Total metals in the IGM}
To address the question of the potentially missing metals, \citet{Danforth2014} have completed the largest IGM survey to date, consisting of 75 AGN sight lines that pass through over 2,500 unique HI absorption systems. They find in a later paper \citep{Shull2014} that the average metallicity of the IGM in the low redshift universe is $\Omega_{\mathrm{Z}} \approx 10^{-5}$.

In comparison, we find $\Omega_{\mathrm{Z}} = 0.2 - 0.9 \times 10^{-5}$, depending on the model, with E2 having, once again, the lowest metallicity for R1 and R3 reheating models. This is a direct result of limiting the supernova velocity in E2 to L* circular velocity and below. E2 does not assume that all of the supernova energy is put into reheating and ejection, so it follows that it would eject less than its counterparts. 

Likewise, for a given ejection model, the cold gas-dependent reheating method (R3) gives the lowest IGM metal density. As discussed above in Section \ref{s:SMF}, R3 has the highest disk mass loading factor for the mass range of interest. This means that reheating the gas takes more of the available energy and so less gas and fewer metals are ejected.

All combinations of that include R1 or R2 are between $\Omega_{\mathrm{Z}} = 0.6 - 0.9 \times 10^{-5}$, and are reasonable fits to the observations. Moreover, our agreement with \citet{Shull2014}'s observations is consistent with no ``missing metals" problem. The values for each model can be found in Table \ref{tab:Ztable}.

\subsubsection{Distribution of Metallicity}
The most commonly quoted value for the metallicity of the IGM is 0.1\Zsun, which comes from hydrosimulations such as \citet{Cen2001}. Measurements of quasar absorption lines range from $-1.8 \le [\mathrm{O}/\mathrm{H}] \le -0.6$ at $z = 0.12$ \citep{Tripp2001} to $-1.93 \le [\mathrm{O}/\mathrm{H}] \le 0.03$ at $z = 0.08 - 0.23$ \citep{Savage2014}. Other measurements have similar ranges \citep[e.g.][]{Tumlinson2011}.

Our maximum metallicities are all approximately solar, with R2E1 having the lowest maximum at \logten(Z/\Zsun) $\sim -0.136$. Since \cite{Savage2014} find an IGM metallicity higher than this, it would appear this model is less likely than the others, although it does not completely rule it out. All models have their most likely metallicities in the range $-1.5 <$ \logten(Z/\Zsun) $< -1$\ (Figure \ref{fig:Z_hist}), with minimums at zero metallicity and maximums at solar. The observed metallicity ranges fall well within the models.

\citet{Lehner2013} find a bimodal distribution in the metallicity in the circumgalactic medium around 28 Lyman limit systems. While they are surveying the gas within the halo and this paper addresses the gas outside, their results are nevertheless indicative that our mixing model (100\% mixing on relatively short timescales) is an oversimplification. The metallicities we find in Figure 8 have slightly lower maximum and significantly lower minimum values, but that is to be expected as gas flows out of the halo.

Since observations are limited to quasar sightlines, measuring the IGM's proximity to a galaxy of a given mass is a rare event. Nevertheless, \citet{Kacprzak2014} find a M$_{\star} = 10^{10.6}$\Msunnh\ ($h = 0.7$) galaxy at $z\sim 0.2$ with a stellar metallicity of [O/H] = -0.21$\pm$0.08 and an IGM metallicity (58 kpc\ away) of [X/H] = -1.12$\pm$0.02. This separation most likely puts the measurement within the hot halo, but the authors determine the gas is infalling and therefore comes from the IGM. Additionally, \citet{Prochaska2011} find OIV out to 300 kpc around sub L* galaxies, well outside of the virial radii. While none of our models fail to reproduce these observations in IGM metallicities, runs which use R2 tend to have higher galaxy metallicities, making them less likely to match \citet{Kacprzak2014}'s galaxy overall.

We would need to tune SAGE to each reheating/ejection combination to absolutely rule out any model, but by comparing the current results to observation, we can identify weaknesses in the models. E2 is based on the logical assumption that not all supernova energy is converted into kinetic energy. Yet it under-predicts the median and total metals in the IGM. For E2 to successfully reproduce observations, there needs to be an additional mechanism that pushes metals into the IGM around sub-L* galaxies. One possibility is that it preferentially pushes relatively high metallicity gas from the halo.

\section{Summary}
\label{s:summary}

In this paper, we have measured the effect of supernova feedback on the contamination of the IGM by ejected metals. To do this, we have tested nine combinations of reheating and ejection models from supernova feedback. The three reheating models assume different physics for the disk mass loading factor: no dependence on the host galaxy, inversely dependent on the circular velocity, and dependent on the distribution of cold gas in the disk. Likewise, the three ejection models assume a constant halo mass loading factor with either a constant or a circular-velocity dependent wind, and a halo mass loading factor that is dependent on the amount of mass that is reheated.

To model the IGM, we have developed a method for rapidly modelling baryons in the diffuse intergalactic medium akin to the semi-analytic models of galaxy formation. It can be used in conjunction with the galaxy models to measure the effects of feedback on a galaxy's \env. By varying the efficiency with which a supernova reheats gas from the disk and then ejects hot gas from the halo, we can make predictions concerning the metallicity of the IGM.

We measure the metallicity of the IGM to be at its highest around galaxies with $10.0 < $\logten(\Mstar/\Msun)$ < 10.5$\ in \env s about $10 \times \bar{\rho}$. The actual value of this maximum is highly dependent on the combination of reheating and ejection models used. Most ($> 90 \%$) of the metals in the IGM are ejected by galaxies with \Mstar $< 10^{10.33}$\Msun\ in \env s with $\rho < 30\times \bar{\rho}$, regardless of the reheating/ejection model used. Our results indicate that the regions around small groups are ideal places to see metals in the IGM compared to both observations and hydrosimulations.

In agreement with both observations and hydrosimulations, we find 5-10$\%$\ of all metals are in the IGM. Our metallicity ranges also compare favourably with observations of the IGM near individual galaxies. 

A key question in current IGM studies is the evolution of its metallicity with redshift. To answer this question, we need to fully couple our galaxy model, SAGE, with the diffuse gas model, SWAMI. This will allow us to leave ejected metals behind as galaxies move through the cells, and will be the subject of future work.

\section*{Acknowledgments}
The authors would like to thank Emma Ryan-Weber for her useful discussions and the referee, Brad Gibson, for his helpful comments. GMS is supported by a Swinburne University SUPRA postgraduate scholarship. DC acknowledges the receipt of a QEII Fellowship from the Australian Research Council. AB is supported by Swinburne APA and IPRS postgraduate scholarships.

\bibliographystyle{mn2e}
\bibliography{manuscript}

\end{document}